\documentclass[12pt]{article}
\usepackage{graphicx}
\usepackage{amssymb,amsmath,amsfonts,palatino,amsthm}
\usepackage{amssymb}
\usepackage{epstopdf}
\usepackage{slashed} 
\newcommand{\f}{\begin{equation}}
\newcommand{\ff}{\end{equation}}
\newcommand{\blankline}{\vskip .3cm}
\begin{document}

\title{Four principles for quantum gravity \\}
\author{Lee Smolin\thanks{lsmolin@perimeterinstitute.ca} 
\\
\\
Perimeter Institute for Theoretical Physics,\\
31 Caroline Street North, Waterloo, Ontario N2J 2Y5, Canada}
\date{\today}
\maketitle

\begin{abstract}
 
Four principles are proposed to underlie the quantum theory of gravity.  We show that these suffice to recover the Einstein equations.  
We also suggest that MOND results from a modification of the classical equivalence principle, due to quantum gravity effects.

\end{abstract}

\newpage

\tableofcontents


\section{Introduction}

{\it This paper is dedicated to Thanu Padmanabhan on his 60th birthday, as it reflects longstanding  concerns and insights 
of his\cite{Paddy1,Paddy2,Paddy3,Paddy5,Samantray-Paddy}.  With thanks for his vibrant originality and dedication to the search for the fundamental principles of nature.  }

\blankline
\blankline

Albert Einstein\cite{AE} taught us to distinguish between {\it principle theories} and {\it constructive theories.}   The latter are descriptions of particular phenomena, fields or particles that constitute nature.  These  are specified in terms of dynamical equations of motion that the constituents obey.    Principle theories are different: they give us universal principles that all physical phenomena must obey, whatever fields or particles constitute nature.   The paradigmatic example Einstein used is the laws of thermodynamics.  Einstein used the distinction to argue that special relativity is a principle theory.  He used this to distinguish special relativity from its rivals, principally the Lorentz theory of the electromagnetic aether, which he argued is a constructive theory.  The lesson is that when one can manage to encompass a phenomena in terms of a principle theory, that that will likely be superior to a constructive theory. 

Once we have a principle theory, we can use it to frame and constrain candidate constructive theories.  

In this contribution I propose we frame quantum gravity as a principle theory.   Loop quantum gravity, string theory, causal sets, CDT, etc can all be seen as constructive theories that tell us what quantum spacetime might be "made of".   These may contain some elements of the truth about what constitutes quantum spacetime.  But I would like to suggest an alternative road in which we first seek principles.  

These principles should have some non-trivial consequences.  First of all, they should reproduce what we know already. In particular, the field equations of general relativity should emerge in a suitably defined classical or coarse grained limit.  In addition, they should entail novel phenomena or provide an unexpected explication of  known phenomena.

In this paper, I propose four principles and show that together they do entail that general relativity emerges as a coarse grained approximation.  We also get new insights into the physics at small cosmological constant, $\Lambda$, both negative and positive.
In the case of $\Lambda < 0$ we find evidence that 
 aspects of the $AdS/CFT$ correspondence are related to these principles.  This is not new, but involves an interpretation of some existing results.  But in  the opposite case of small, positive, cosmological constant, I find a surprise:  a tentative argument that the principle of equivalence, and hence gravity, is modified at low accelerations, compared to 
\f
a_* = c^2 \sqrt{\frac{\Lambda}{3}}.
\ff  
As I discuss in section 5, this may be related to MOND.


Here are the four  principles:

\begin{enumerate}

\item{}{\bf The principle of absolute causality and relative locality}  {\it A quantum spacetime consists of a set of events whose fundamental properties include causality (i.e. causal relations), energy and momentum.\cite{CS,ECS1,ECS2,ECS3,Cohl}\footnote{The notion of causal sets was introduced by \cite{CS}.  Causal sets built out of intrinsic structures was developed by \cite{Cohl,CS2,FotiniCS}.}.
Classical spacetime is emergent, as is locality.  Locality is also relative to the positions of observers and the energy and other properties of the probes they employ to measure and tracks distant causal processes\cite{RL1,RL2}.}

\item{}{\bf Correspondence principle }  {\it Classical spacetimes emerge in a suitable limit, by coarse graining the causal structure
of sufficiently large quantum spacetimes.}

\item{}{\bf The weak holographic principle:}  {\it The area of a surface, which is defined by the causal structure as the boundary of  a subsystem, is a measure of the channel capacity of that surface to serve as a channel of information in and out of that subsystem.\cite{weakholo}. } 

\item{}{\bf  The quantum equivalence principle: }{\it  The observers who, in the absence of curvature and a cosmological constant, see 
the vacuum to be a maximal entropy thermal state, are those that in the classical limit are uniformly accelerating.  Hence those observers who see the vacuum to have zero temperature must be inertial\cite{fluctuations}. }

\end{enumerate}

The first is a strengthing of relative locality as originally proposed in \cite{RL1,RL2}\footnote{A different proposal for relative locality is in \cite{RLM}.}. We should stress that the correspondence principle guarantees the existence of a classical manifold and metric,  $g_{ab}$, but does not require that that metric satisfy the Einstein equations or any other dynamical law.  

The third principle is not new\cite{tHooft-holo,Lenny-holo,weakholo}, but it will be important to propose a form of it suitable for a fundamental theory of quantum gravity.  The last is a development of a principle proposed in \cite{fluctuations}, and will require some explanation.

I should note that it may be attractive to think that the holographic principle is a consequence of the others, particularly the quantum equivalence principle.  But there is a simple  reason to think the holographic hypothesis and the quantum equivalence principle  are independent.  This is that they introduce independent constants.  The holographic hypothesis introduces a quantum of area, $A_p$, to define surface entropy.  The quantum equivalence principle introduces $\hbar$, in the guise of a boost temperature (with dimensions of the boost Hamiltonian, which are action.)  The ratio gives us Newton's constant, $G=\frac{A_p}{\hbar}$.  We will see below how they work together to give us Einstein gravity.

The quantum equivalence principle does more than deepen the classical equivalence principle, it puts limits on the older 
idea.  These occur in the presence of a positive cosmological constant, which arises because there is no observer that sees the vacuum to have zero temperature.  I argue in section 5 that this leads to a violation of the equality of gravitational and inertial mass.  This violation is essentially a 
renormalization group effect, hence it depends on temperature.  But because of the quantum equivalence principle, this temperature dependence transmutes into an acceleration dependence.  This, I briefly show in section 5, can explain
MOND\cite{MOND,MLS}, and hence obviates the need for dark matter to explain the 
galaxy rotation curves\footnote{This does not yet address the need for dark matter on scales of clusters and large scale structure.  It is possible that these are explained by dark matter while MOND explains the galaxy rotation curves.}.


Once we have a set of principles from which the semiclassical limit follows, we can seek to show that various constructive theories realize the principles.  For example, a key problem with background independent approaches to quantum spacetime histories, such as spin foam models, causal sets and causal dynamical triangulations, is showing that they have a good classical limit. This problem has two aspects.  First, we have to show that there is a classical limit described in terms of an emergent classical spacetime.  Then one has to show that the metric of this spacetime satisfies the Einstein equations.  Rather than approach these questions directly,  the principles I propose guarantee these outcomes, the first directly as a result of the correspondence principle, the second as a consequence of them all.  

An objection that might be made of our strategy is that loop quantum gravity (or string theory, or CDE) is already a principle theory, whose principles are merely those of general relativity and quantum field theory.  That is partly true, but it is important to mention that along the way from principles to the physics there are, in each case, technical choices that need to be made.  These choices make the theory at least partly constructive.   We are interested in a different question which is whether there are any new principles acting in the world which govern quantum gravitational phenomena.


I begin introducing the new principles in section 2, 
The aim of section 3 is to sketch the recovery of general relativity in a suitable limit.  Section 4 contains a short discussion of how the $AdS/CFT$ correspondence is related to the quantum equivalence principle proposed here.  In section 5, I sketch a route to deriving the phenomenology of MOND from the quantum equivalence principle, after which we conclude.

\section{Four principles for quantum gravity}

I now propose four principles for quantum gravity.  In each case there is a short statement of the principle, followed by a more detailed explication.

\subsection{The principle of absolute causality and relative locality}

{\bf The principle of absolute causality and relative locality.} {\it Causal relations are fundamental to nature as are energy and momentum, whose flow follows those relations. Classical spacetime is emergent, as is locality.  
 Locality is also relative to the positions of observers and the energy and other properties of the probes they employ to measure and tracks distant causal processes\cite{RL1,RL2}. }

\blankline
\blankline

{\bf Explication:} {\it This first principle tells us what the theory is to be about.  We want to specify that, just as in general relativity, a quantum spacetime can be understood to be about events and their causal relations.  The principle also 
asserts the primacy of causality and of energy and momentum over spacetime and locality.

A quantum spacetime consists of a set of events whose fundamental properties include causality (i.e. causal relations), energy and momentum.\cite{CS,ECS1,ECS2,ECS3,Cohl}\footnote{The notion of causal sets was introduced by \cite{CS}.  Causal sets built out of intrinsic structures was developed by \cite{Cohl,CS2,FotiniCS}.}.
 Further, for every causal diamond ${\cal CD}(e,f)$ in the quantum spacetime, we posit that there corresponds a 
Hilbert space,  ${\cal H}_(e,f)$ which records quantum information measurable on the waist, ${\cal B}(e,f)$. }  

There are a variety of structures which exist such as causal pasts, causal futures, causal diamonds and their boundaries.  

We briefly recall their definitions.

\begin{itemize}

\item{}{Causal diamonds on causal sets}

We are given 
two distinct events $e < f$ in a causal set.  We define immediately the causal past of $f$, denoted ${\cal P}(f)$, consisting of $d$ such that $d<f$.  The immediate causal past of $f$ is the subset of ${\cal P}(f)$ consisting of those events reached from $f$ by one causal link into the past, and is denoted, ${\cal IP}(f)$.  Similarly we define the causal future,  and the immediate causal future,  of $e$, denoted respectively by  ${\cal F}(e)$ and  ${\cal IF}(e)$.  The causal diamond is 
\f
{\cal CD}(e,f)= {\cal F}(e) \cap {\cal P}(f)
\ff
The boundary of the past of $f$, denoted $\partial {\cal P}(f)$ consists of elements of ${\cal P}(f)$ some of whose immediate futures are not containned in ${\cal P}(f)$.  Similarly, we define the boundary of the future of $e$, denoted
 $\partial {\cal F}(e)$.  The {\it waist} of the causal diamond, ${\cal CD}(e,f)$, is their intersection.
 \f
 {\cal W}(e,f)= \partial {\cal F}(e) \cap \partial {\cal P}(f)
 \ff

Quantum mechanics is fundamentally about one subsystem of nature 
probing the rest\footnote{This idea is developed in relational quantum theory\cite{louisRQM,meRQM,CarloRQM} and relative locality\cite{RL1,RL2}.}.  The most elementary act of observation a subsystem of the universe can make is to send a probe out into the world at one event and receive a response back at a future event.  The act of probing the world is represented by a causal diamond, making them primary structures.  
Furthermore, the geometry of momentum space is more fundamental than the geometry of spacetime.

  \end{itemize}
  


\subsection{The correspondence principle}

{\bf The correspondence principle:}  {\it Classical spacetimes emerge in a suitable limit, by coarse graining the causal structure
of sufficiently large quantum spacetimes.}

\blankline
\blankline

{\bf Explication:}  {\it Consider a quantum spacetime, ${\cal Q}$, a subset of whose causal diamonds have large spacetime volumes (In Planck units.)  To it there corresponds a classical metric spacetime, $({\cal M}, g_{ab})$ such that for every causal diamond of ${\cal Q}$ whose spacetime volume 
${\cal V}(e,f)$ is sufficiently large, there is a corresponding causal diamond, $\tilde{\cal CD}(e,f)$,  in  $({\cal M}, g_{ab})$,
whose spacetime volume and waist area coincide.}

Hence, among the observables in ${\cal H}_(e,f)$ is the area of its waist, $\hat{\cal A }[{\cal B}]$,
 the spacetime volume, ${\cal V}(e,f)$, and the
spacetime curvature scalar averaged over the causal diamond, given by 
$<{\cal R} >_(e,f)$.  Moreover, we require that there is an action of the Lorentz group on  ${\cal H}_(e,f)$, denoted ${\cal L}.$.
The correspondence allows us to pull back $\cal L$ to the action of $SL(2,C)$ on functions on $S^2$.

The waist of the classical causal diamond is an $S^2$.  There is a mapping from functions, $f$, on the $S^2$ into states
$\psi (f)$ in  ${\cal H}_(e,f)$.  

 We call the pair, quantum and  classical, a {\it paired causal diamond.}
 
Note that in the context of a constructive approach to quantum gravity, such as causal sets, $CDT$ or spin foam models, this is a result we seek to demonstrate.  Whether a given constructive quantum gravity theory satisfies the correspondence principle is then a test for adequacy of a quantum gravity theory.  Thus, this correspondence principle plays the same role as the correspondence principle did in the development of the quantum theory, it is a criterion for adequacy, which the Schroedinger quantum mechanics passed by virtue of Ehrenfest's theorem.  

We should also expect that there are, as in quantum physics, limits to the correspondence principle that arise from novel phenomena.  In the present case these could arise from relative locality as well as from the ultra-low acceleration $MOND$ regime.  It is interesting that both point to modifications of general relativity at large distances.

\subsection{The weak holographic  principle}


{\bf The weak holographic principle:}  {\it The area of a surface, which is defined by the causal structure as the boundary of  a subsystem, is a measure of the channel capacity of that surface, to serve as a channel of information in and out of that subsystem.\cite{weakholo}. }

\blankline
\blankline

{\bf Explication:}  {\it In a quantum causal structure we can define special surfaces as the intersection of causal past of one event with the causal future of another. The area of these surfaces are a measure of the
channel capacity of these surfaces.  In more detail, let us consider a quantum causal diamond, ${\cal CD}(e,f)$, to which there corresponds a classical causal diamond $\tilde{\cal CD}(e,f)$.  A consequence of the metric geometry is that the waist or corner, $\tilde{\cal B}$ of the classical causal diamond is a space like $S^2$.  By the correspondence to the quantum causal structure,  
$\tilde{\cal B}$ inherits the Hilbert space, ${\cal H}(e,f)$ and its observables.  These include the
area $ \hat{\cal A }[{\cal B}]$.   The Hilbert space ${\cal H}(e,f)$ represents the information that the observer represented by a causal diamond may obtain about the world by means of probes that measure information coming into its waist.  



\f
S[{\cal B}] = \frac{\cal A [{\cal B}]}{A_p}
\ff

We will find below that $A_p = 4 \hbar G$,  but it is introduced as an independent constant with units of area.}

\subsection{The quantum equivalence principle}


{\bf  The quantum equivalence principle: }{\it  The observers who, in the absence of curvature and a cosmological constant, see 
the vacuum to be a maximal entropy thermal state, are those that in the classical limit are uniformly accelerating.  Hence those observers who see the vacuum to have zero temperature must be inertial\cite{fluctuations}. }

\blankline
\blankline

{\bf Explication:}{\it  Worked out in detail, the principle has two parts.  The first part is purely quantum, the second develops the correspondence between the quantum and classical aspects of a paired causal diamond.}


The first part specifies the existence of dimensionally reduced thermal states corresponding to uniformly accelerated observers, boosted to the infinite momentum frame.

{\it Consider an observer in a quantum spacetime who is uniformly accelerated to an arbitrarily high boost.  The observer sees physics to be conformally invariant, dimensionally reduced by the elimination of the longitudinal direction, and thermal in the sense that it is described by the maximally entangled state generated by the boost Hamiltonian.   

To every quantum causal diamond, in the quantum causal structure, and every boost generator, $\hat{K}_z$  in ${\cal  L}$ we associate a thermal state, with maximal entanglement entropy,
in a $2+1$ (generally a $(d-1)+1$) dimensional conformal field theory, which lives on the waist, 
This is given by\cite{modularH}
\f
\rho_U = e^{-\frac{2 \pi }{\hbar}H_{Boost}}
\label{mod}
\ff
where $H_{Boost}$ is the positive definite boost Hamiltonian, corresponding to  $\hat{K}_z$ 
which acts on degrees of freedom on the waist.  Note that the boost Hamiltonian must be dimensionless.  The dimensionless Unruh temperature is\cite{Unruh1},}
\f
T_U =\frac{\hbar}{2 \pi } 
\label{TU}
\ff


The second part requires that when the classical spacetime emerges, the locally accelerated trajectories correspond to observers that describe the vacuum as a thermal state with maximal entanglement entropy, as was posited in \cite{fluctuations}.

Specifically, it requires that
{\it  in the cases in which  ${\cal CD}(e,f)$ is a paired causal diamond, there corresponds to it a causal diamond in a classical spacetime.  The  classical metric $g_{ab}$ has a radius of curvature, $R$\cite{Ted2015}.  The state (\ref{mod}) corresponds to what an observer  in the dual classical spacetime, uniformly boosted to the infinite momentum frame, might see.  More specifically,
in the limit where the  spacetime volume is large in Planck units, but small compared to $R^4$, the metric has an approximate boost killing field. Then, \it $\rho_U$ describes the quantum state of the causal diamond as seen by an observer 
	moving with respect to an approximate boost killing field of the metric $g_{ab}$.}

\subsubsection{Motivation:}  

In the classical equivalence principle we make use of the notion of a freely falling, or inertial, frame, which the Lorerntzian geometry gives us before hand.  We then define a uniformly accelerated frame as one which is accelerated relative to an inertial frame.  

But the notion of an inertial frame does not appear naturally in the fundamentals of quantum gravity.  The reason is that it depends on a  limit of weak coupling to define.  An inertial frame is one that {\it ``has no forces on it."}  Except, 
of course, that it must interact with its surroundings if it is to serve as an observer.  Such an observer can only make sense when defined in a weak coupling limit.  
However, such a limit is antithetical to being deep in the quantum gravity regime, where all the degrees of freedom are interacting with each other. 

Thus, in the quantum version we start with the accelerated elevator and later take a weak coupling limit where we transform to the freely falling frames.

What we want, then, is a quantum notion of a uniformly accelerating frame.  
The result of Unruh suggests that this should correspond to a thermal state. Notice that uniform acceleration implies an unlimited boost,  corresponding to a uniform acceleration carried out for an unlimited time.  In this time, an observer is accelerated to a boost, $\gamma$, relative to its initial motion, which is arbitrarily large.  Thus we need a notion of an
observer corresponding to what high energy physicists call the {\it infinite momentum frame\cite{IMF}.}   The quantum equivalence principle thus is going to assert that the limit of a uniformly accelerating observer is related to the infinite momentum frame.

We recall that in the infinite momentum frame the longitudinal coordinate is length contracted to an arbitrarily small interval, so that spheres are quashed down into pancakes.   Thus a uniformly accelerating observer should see a world reduced by the elimination of the longitudinal direction.   

Note that the limit of infinite boost, when $\gamma \rightarrow \infty$, can be seen as the limit of a renormalization group transformation.  Asking that the quantum physics have a limit is then the same as asking that there be a fixed point of the renormalization group, i.e. a scale invariant theory.  So the limit theory has to be a conformal field theory.  

We can say this a different way. Greenberger\cite{Greenberger} explains why, strictly speaking,  the  equivalence principle cannot hold for a particle of finite mass.  The fact that the Compton wavelength $\lambda_C = \frac{\hbar}{m}$ is finite means that how a massive particle falls in a gravitational field will depend on its mass.   But this does not rule out applying the equivalence principle to massless particles.   If we generalize this to QFT we would say that any mass scale in a QFT is an impediment to the satisfaction of the equivalence principle, hence the quantum equivalence principle describes what a uniformly boosted observer sees in terms of a conformal field theory.  

Moreover, the conformal field theory we need is one whose degrees of freedom can be attributed to the waist, which is a two dimensional sphere.  Hence the longitudinal direction disappears.  This is a well known characteristic of physics in the infinite momentum frame.   If an object of longitudinal size, $r$ is length contracted by a 
$\gamma$ so large that $\frac{r}{\gamma} < l_{Pl}$ then it no longer is describable in terms of extension in a classical geometry.  Either it disappears because it becomes part of quantum geometry at the Planck scale (if  the lorentz symmetry is unmodified) or it disappears because it gets squeezed down to $l_{Pl}$ as a limit in a  deformed version of lorentz symmetry.

We note that this disappearance of a dimension in a short distance of high energy limit may be connected to the phenomena of dimensional reduction observed in several approaches to quantum gravity\cite{fractal,reduction}.

\subsubsection{Comments on the quantum equivalence principle}

We can make some simple comments on this new proposal:

\begin{itemize}

\item{}{\bf Note 0:} Equation (\ref{TU}) is where $\hbar$ is introduced. 

\item{} {\bf NOTE 1:} The quantum equivalence principle incorporates the Unruh effect\cite{Unruh1}.

\item{} {\bf NOTE 2:} We can conjecture that the $AdS/CFT$ correspondence is at least partly a consequence of the quantum equivalence principle, combined with the weak holographic principle,  because as you go to the boundary of $AdS$ the radius of curvature goes to a large constant.   We discuss this below.


\item{}  {\bf NOTE 3:} The quantum equivalence principle incorporates the {\it equivalence of quantum and thermal fluctuations} posited by myself\cite{fluctuations} and developed by Kolekar and Padmanabhan\cite{KP}.  
In that paper, \cite{fluctuations}, I raised the question of why the class of special observers that see zero temperature in the ground state are the same as the special class of inertial observers.  This is suggested by the observation that they need not be the same, because the former depend on the choice of vacuum in the $QFT$, while the latter does not.

Coming from classical physics we consider the class of inertial observers as prior to and more fundamental than, the class of zero temperature observers.  In quantum gravity, we must turn this around.  In the deep quantum gravity regime, governed by the quantum equivalence principle, there are no particle trajectories and no notion  of inertia or acceleration.  Moreover, associated to causal diamonds, there is a dimensionless notion of boost energy and boost temperature.  So the fundamental notion is the dimensionless temperature, 
$T_U$, from which the dimensional temperature and corresponding acceleration emerge in the appropriate limit following the breaking of conformal invariance.

\item{}  {\bf NOTE 4:} The quantum equivalence principle (QEP) implies the classical equivalence principle (CEP), in the limit $\hbar \rightarrow 0 $ and in the absence of a cosmological constant.  The dimensional reduction requires that the acceleration proceed to a boost that exceeds. for any physical length scale, $L$, 
\f
L^\prime = \frac{L}{\gamma} < l_{Pl}
\ff
but in the limit $\hbar \rightarrow 0 $, $l_{Pl} \rightarrow 0$ so the condition is never met.  Furthermore the Unruh temperature also goes to zero.  

In the presence of a positive cosmological constant, the limit to the classical equivalence principle may be modified for
small accelerations, $a < a^*$, where $a^* = c^2 \sqrt{\Lambda}$.  This is discussed in section 5, below.

\item{}{\bf NOTE 5}:  The requirement that $\rho_U$ be an equilibrium state implies that it is the maximal possible entropy.  But that is limited by the channel capacity.  Hence we have
\f
S_B = - Tr \rho \ln \rho = \frac{1}{T_U} < H_B > 
\ff

\end{itemize}

\subsubsection{Free fall}

The classical equivalence principle has two parts, related to accelerating elevators and elevators in free fall.  We started
with the analogue of accelerating elevators, can we get back to a statement of a quantum equivalence principle
for freely falling observers?

The key idea is that $\rho_U$ is the maximal entropy state.  A freely falling reference frame is one that will observe the minimum entropy state, which is the vacuum.  This means that the notion of inertial motion, i.e free fall, should be a consequence of the possibility of reducing entanglement entropy to a minimum by transforming from a thermal state to the vacuum, below the scale of the radius of curvature.  This is essentially the hypothesis made in \cite{fluctuations} which posits that it cannot be contingent or coincidence that observers who see minimal entanglement entropy move inertially.

To do that we have to break the conformal invariance that got us to the limit of large boosts, which made it possible to use the boost Hamiltonian and boost temperature, both of which have units of action rather than energy.  The breaking of conformal invariance gives us a  time scale, $\tau $.  We 
can then define the dimensional boost Hamiltonian and temperature,
\f
\tilde{H}_{B} = \frac{H_{Boost}}{\tau}, \ \ \ \ \ \ \tilde{T}= \frac{\hbar}{2\pi \tau}= \frac{\hbar a }{2\pi c }
\label{renH}
\ff
where the acceleration $a= \frac{c}{\tau}$.  The thermal state is then
\f
\rho_U = e^{-\frac{2 \pi c}{\hbar a }\tilde{H}_{B}}
\ff
In the absence of a positive cosmological constant, we then take the limit of $a \rightarrow 0$ to find the minimal entropy, or ground, state,
\f
\rho_0 = \lim_{a \rightarrow 0} e^{-\frac{2 \pi c}{\hbar a }\tilde{H}_{B}}
\ff

If this state exists it will have at least approximate symmetries which generate the symmetries of spacetime, and hence lead to the recovery of the equivalence principle for freely falling observers.

The fact that the ground state has translation symmetry implies that the specification of which motions are inertial does not depend on any property of a freely falling particle.  This implies the equality of gravitational and inertial mass,
\f
m_I = m_g.
\ff

In section 5, we will see how this is modified when the cosmological constant is positive.

\section{Recovery of the Einstein equations}

We show that these principles suffice to recover the Einstein equations.  We follow a strategy pioneered by 
Jacobson\cite{Ted95,Ted2015} and Padmanabhan\cite{Paddy5}.

Suppose that we have a boosted observer inside a paired causal diamond.  
By the weak holographic principle, the entropy is 
\f
S_B = \frac{A(B)}{A_p}
\ff

By the quantum equivalence principle, the entropy
is also 
\f
S_B = - Tr \rho \ln \rho = \frac{1}{T_U} < H_B > 
\ff
This is of course the first law of thermodynamics, emerging here as a consequence of the weak holographic principle and the quantum equivalence principle.

It follows that
\f
<H_B > = \frac{1}{8 \pi G} A[B]
\ff
where $G=\frac{A_p}{4\hbar}$.  

This has been called the first law of classical spacetimes.  

Now consider three events, $A < B < C$ which generate two paired causal diamonds ${\cal CD}_1 ={\cal CD}[A,B]$ which is a subset of 
${\cal CD}_2 ={\cal CD}[A,C]$.  Then we can also show that
\f
<\Delta H_B > = <H_{B_2} > -<H_{B_1}  > =  \frac{1}{8 \pi G} \Delta A[B] = \frac{1}{8 \pi G} \left [ A[B_2] - A[ B_1] \right ]
\label{DeltaHB}
\ff


We next use the correspondence principle to describe the boundary of the causal diamond in terms of a congruence of null geodesics in the emergent spacetime.  This is specified by an affine parameter $\lambda$ and a null tangent vector $k^a$.
We express $\Delta A[B]$ in terms of the expansion $\theta$.
\f
\delta A= \int dA d\lambda \theta
\ff
To compute this we will use the Raychauduri equation for a  congruence of null geodesics
\f
\dot{\theta} = - \frac{1}{2} \theta^2 - 2 \sigma^2 + 2 \omega^2 - R_{ab}k^a k^b
\ff

We next note that every event $e$ in a causal spacetime appears in the corners of many causal diamonds.  We fix an arbitrary event and look for suitable causal diamonds.  Suitable means that $e$ is in the corner of the causal diamond and that the null geodesics that make up the boundary of that causal diamond satisfy that $2 \sigma^2$ and  $\omega^2$ are negligible compared to $R_{ab}k^a k^b$.
\f
 \sigma^2, \omega^2 << R_{ab}k^a k^b
\label{it}
\ff

We also assume that an event $e$ in the corner, $B_1$ can be chosen such that there
\f
\theta_1 << R_{ab}k^a k^b
\label{lt2}
\ff
If this is not the case, but (\ref{it}) are satisfied we just wait, and consider an event $e^\prime$ further along the light cone.  This is because under our assumptions.
\f
\dot{\theta} = - \frac{1}{2} \theta^2  \rightarrow   \theta \approx \frac{2}{t}
\ff
so that after some time (\ref{lt2}) will be satisfied.   When it is we have
\f
\delta A= \int dA d\lambda \theta = \int dA d\lambda \lambda dot{\theta} = -  \int dA d\lambda \lambda R_{ab} k^a k^b  
\ff

Now we work on the other side of (\ref{DeltaHB}).  The assumption that the shear $\sigma$ can be neglected means that there is not a lot of energy in gravitational radiation.  Hence this implies that
\f
\Delta H_B = \Delta Q = \int_{\cal H} T_{ab} \xi^a d\Sigma^b
\ff
where $xi^a$ is an approximate killing field generating the boost near the corner.  By the equivalence principle this must exist.  We can  follow Jacobson\cite{Ted95} in setting
\f
\xi^a = - \kappa \lambda k^a
\ff
\f
d \Sigma^a = k^ad\lambda dA
\ff
hence we have
\f
\Delta H_B  = \int_{\cal H} T_{ab} k^a k^b \lambda d\lambda dA
\ff
We now use the fact that (\ref{DeltaHB}) will be true for a large number of causal diamonds whose waist includes $e$.  We then have
\f
R_{ab} + g_{ab} f = T_{ab}
\ff
Making use of the Bianchi identities we have
\f
R_{ab}- \frac{1}{2} g_{ab} R - g_{ab}\Lambda = 8 \pi G T_{ab}
\ff

Thus we see that our principles imply that general relativity is satisfied.


\section{AdS/CFT as an example of the quantum equivalence principle}

The formulation of the quantum equivalence principle is inspired by early ideas of holography and the infinite momentum frame, a connection that has been emphasized by Susskind\cite{Lenny-bitstring,Lenny-holo}.  However, there are interesting
implications of the principle in the case of negative cosmological constant, which appear to be related to aspects of  
the $AdS/CFT$
correspondence.

Notice that, as pointed out by\cite{DL1,RindlerQG,holegraphic,Parikh-Samantray,Samantray-Paddy}, an observer at large $r >>R$ of a Schwarzschild-$AdS$ spacetime is in a situation analogous to a uniformly accelerated observer in flat spacetime, i.e. an observer in Rindler spacetime.
The uniformly accelerated observer in the $IMF$ limit sees all massive particles moving with respect to it in the negative direction.  She sees particles initially ahead of her lose speed and then fall behind, as she passes them.

Similarly, all outgoing massive particles in asymptotically $AdS$ spacetimes reach a maximal $r$ and then fall back.
The uniformly accelerating observer sees light coming from behind it to be increasingly redshifted, where that redshift goes to infinity in the $IMF$ limit.

To see this we work in global coordinates\cite{DL1}.
\f
ds^2 = - f(r) d\tau^2 + \frac{dr^2}{f(r)} + r^2 d\Omega^2
\label{global}
\ff
where $f(r)= (1-\frac{2GM}{r} + \frac{r^2}{R^2} )$,  with $\Lambda =-\frac{1}{R^2}$.  The redshift for outgoing light is
\f
\frac{\omega (r_2)}{\omega (r_1)} = \frac{\sqrt{f(r_1)}}{\sqrt{f(r_2)}} = 
\sqrt{\frac{1-\frac{2GM}{r} + \frac{r^2_1}{R^2} }{1-\frac{2GM}{r} + \frac{r^2_2}{R^2} }}
\ff
In the limit $r_2 > r_1 >> R > GM$ this is 
\f
\frac{\omega (r_2)}{\omega (r_1)} =  \frac{r_1}{r_2}
\ff
This is the same as in Rindler spacetime where $f(r) = \frac{r^2}{R^2}$.   In the limit that $r_2 \rightarrow \infty$ the redshift factor
goes to zero.  This has been seen as a manifestation of an $IR/UV$ duality in which high frequency excitations in the bulk are redshifted to zero energy and infinite wavelength by the time they reach infinity.

It has also been found by Deser and Levine\cite{DL1} that a uniformly accelerating observer with uniform acceleration $a$, in anti-deSitter spacetime observes a temperature
\f
T= \frac{\sqrt{a^2 + \frac{\Lambda}{3}}}{2 \pi c} = \frac{a_5}{2\pi c } =  \frac{\sqrt{a^2 - \frac{1 }{R^2 }}}{2 \pi c} 
\label{T1}
\ff
where $a_5$ is the acceleration of the worldline in a five dimensional Minkowski spacetime in which the $AdS$  spacetime is embedded.  

We note that for the temperature to be non-zero,  the acceleration must satisfy $a > \frac{1}{R}$.  This is required for the accelerated observer's worldline to have an horizon.   This is  {\it not} the case for an observer at constant $r$ in the global coordinates (\ref{global}), instead one can show that for all such observers $a < \frac{1}{R}$.  

This requires modifications of the quantum equivalence principle, because there are observers with non-zero acceleration, but vanishing temperature.

Observers with $a > \frac{1}{R}$ do exist, some of them can be described as the constant $\xi$ worldlines, in a coordinate system in which $\xi$ replaces $r$ and in which  the $AdS$ metric is expressed as 
\f
ds^2 = -\frac{\xi^2}{R^2} dt^2 +\frac{d \xi^2}{1+ \frac{\xi^2}{R^2}}+ (1+ \frac{\xi^2}{R^2})[d\chi^2 + R^2 \sinh^2 (\frac{\xi}{R}) d\Omega^2_{d-2}]
\ff

Note that $\xi=0$ is an horizon for the accelerated observers at constant $\xi >0$.  These have accelerations 
\f
a^2 = \frac{1}{R^2} + \frac{1}{\xi^2}
\ff
which are all greater than $\frac{1}{R^2}$, hence by (\ref{T1}) they have non-zero temperatures
\f
T= \frac{\hbar}{2\pi \xi}
\ff

Because $\xi=0$ is an horizon these coordinates cover only a wedge of $AdS$ spacetime, analogous to the way in which Rindler coordinates cover only a wedge of Minkowski spacetime.  Consequently an observer who  can only observe this wedge sees a thermal state.

We note that in the limit $R\rightarrow \infty$ for fixed $\xi$ (or $\frac{\xi}{R} \rightarrow 0$), the $AdS$ metric becomes the pure Rindler metric
\f
ds^2 = -\frac{\xi^2}{R^2} dt^2 +{d \xi^2} +d\chi^2 + \xi^2  d\Omega^2_{d-2}
\ff
Thus, under these conditions the requirements of the quantum equivalence principle are satisfied.  Hence we predict that for large $\xi$, in the presence of larger $R$, the physics is represented by a hot $CFT$ on flat Minkowski spacetime.

This is indeed the case, as has been described in detail in a number of papers\cite{DL1,RindlerQG,holegraphic,Parikh-Samantray,Samantray-Paddy}.

\section{The origin of MOND from the quantum equivalence principle}

We next ask how the correspondences we described in section 2.4 are modified by the presence of a positive cosmological constant, $\Lambda$.
This bathes the system in a bath of low temperature horizon radiation.
It is natural to hypothesize that this results in a temperature dependent renormalization of the scale $\tau$.
\f
\tau \rightarrow \tau^\prime = \tau G^{-1} (\frac{T}{T^*})
\ff
where $G$ is an adjustment of the renormalization scale and $T^*$ is the temperature associated to the cosmological constant scale.
\f
T^*=\frac{\hbar c}{2 \pi } \sqrt{\frac{\Lambda}{3}}
\ff
We would like to compute $G (\frac{T}{T^*})$ from first principles, but below
we will estimate it empirically.

It follows that  the effective Hamiltonian, $\tilde{H}_{B}$ defined by (\ref{renH})  is renormalized
\f
\tilde{H}_{B} \rightarrow \tilde{H}_{B}^\prime =  G (\frac{T}{T^*}) \tilde{H}_{B}
\ff

In the non relativistic limit the Hamiltonian relevant for   a star in orbit in a galaxy has the form of a sum of terms
\f
\tilde{H}_{B} \rightarrow H_{NR} =  \frac{p^2}{2m_i} - m_g U_{New}
\ff
where $m_i$ and $m_g$ are, respectively, the inertial and passive gravitational mass.
These constants can absorb the renormalization factors
\f
m_i^{ren} = \frac{m_i}{G (\frac{T}{T^*}) },  \ \ \ \ \ m_g^{ren} = m_g G (\frac{T}{T^*}) .
\ff
  Hence,
the ratio of gravitational and inertial mass then suffer a temperature dependent renormalization
\f
\frac{m_g^{ren} }{m_i^{ren} } = G^2 (\frac{T}{T^*})
\ff
This modifies the classical equivalence principle.

But as we argued above and in \cite{fluctuations}, the quantum equivalence principle requires that temperature and acceleration be intimately related, by the equivalence of free fall
observers with observers that see minimal entanglement entropy.  So, this turns
into an acceleration dependence
\f
\frac{m_g^{ren} }{m_i^{ren} } = G^2 (\frac{a_{obs}}{a^*})
\ff
where $a^*$ is the acceleration associated to the cosmological constant
\f
a^*= c^2 \sqrt{\frac{\Lambda}{3}}
\ff
and $a_{obs}$ is the observed acceleration of a particle.  
Hence we conclude there must be an acceleration dependent 
modification of the equality of gravitational and inertial mass.  This will be important for extremely
small accelerations, given by the scale of the cosmological constant.

But we also know that
\f
\frac{m_g^{ren} }{m_i^{ren} } = \frac{a_{obs}}{a_{N}}
\ff
which is the ratio of the measured radial acceleration
\f
a_{obs}= \frac{v^2}{r}
\label{v2r}
\ff
to the acceleration predicted by Newtonian theory
\f
a_{N}= \nabla^i U
\ff
Hence we have
\f
a_N = a_{obs}  G^{-2} (\frac{a_{obs}}{a^*})
\ff
We can invert this to find a function $F^2 (\frac{a_{N}}{a^*})$ such that
\f
\frac{a_{obs}}{a_{N}}= F^2  (\frac{a_{N}}{a^*})
\label{ratio}
\ff
This relation, for some function, $ F  (\frac{a_{N}}{a^*})$, is then a consequence of the quantum equivalence principle.

In a recent paper, McGaugh, Lelli and Schubert (MLS) report\cite{MLS} strong confirmation  of an empirical relation of 
this form, first proposed by Milgrom\cite{MOND}.
They measure
$F(\frac{a_N}{a^*})$ in a survey of rotation curves of $153$ galaxies in the SPARC data base\cite{SPARC}.  They measure 
$a_{obs}$, the actual radial acceleration by (\ref{v2r})
at $2693$ radii on these rotation curves.  At the same radii they estimate the Newtonian gravitational potential from baryons as observed in gas and dust, and so determine $a_N $.
They discover that the data is well described by a simple empirical relation of the form of (\ref{ratio}), as shown
in Figure 1.   As they note, it is amazing that such a relation exists over a wide range of galaxy types, sizes and morphologies, as this represents the observed accelerations only by a function of the Newtonian accelerations due to baryons.  
\begin{figure}[t!]
\centering
\includegraphics[width= 0.5\textwidth]{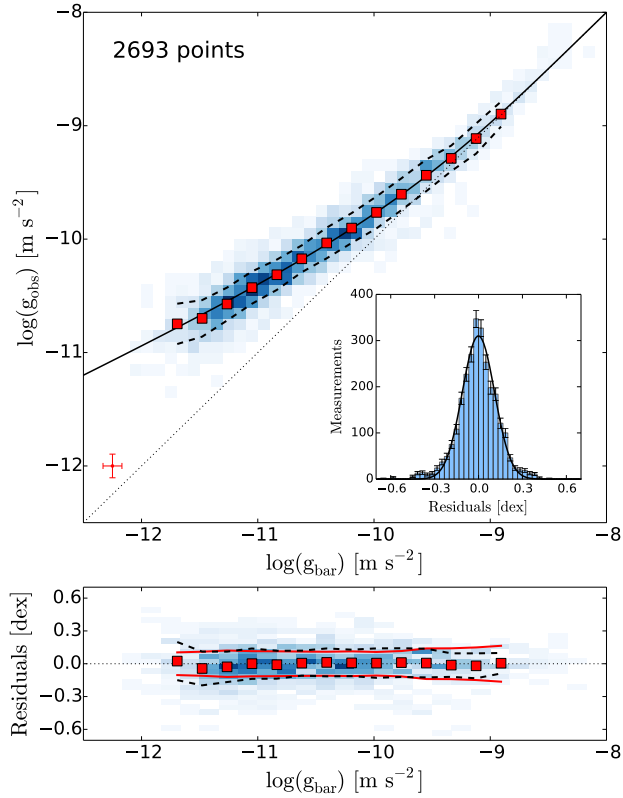}
\caption{The empirical radial acceleration relation, as shown in Figure 3 of \cite{MLS}.  Data is taken from the SPARC data base 
\cite{SPARC}.  Used with permission.  From the original caption:
{\it ``The centripetal acceleration observed in rotation curves, $a_{obs}= g_{obs} = v^2/R$, is plotted against that predicted for the observed distribution of baryons, $a_N=g_b$ in the upper panel. Nearly 2700 individual data points for 153 SPARC galaxies are shown in grayscale. The mean uncertainty on individual points is illustrated in the lower left corner. Large squares show the mean of binned data. Dashed lines show the width of the ridge as measured by the rms in each bin. The dotted line is the line of unity. The solid line is the fit of eq. (\ref{function}),to the unbinned data using an orthogonal- distance-regression algorithm that considers errors on both variables.}  The caption goes on to say,  {\it The inset shows the histogram of all residuals and a Gaussian of width 
$\sigma = 0.11 dex$. The residuals are shown as a function of $g_{obs}$ in the lower panel. The error bars on the binned data are smaller than the size of the points. The solid lines show the scatter expected from observational uncertainties and galaxy to galaxy variation in the stellar mass-to-light ratio. This extrinsic scatter closely follows the observed rms scatter (dashed lines): the data are consistent with negligible intrinsic scatter.\cite{MLS}"}}
\label{data} 
\end{figure}

Furthermore MLS are able to
fit a simple form for $F(a)$ to the data which is\cite{MLS,MS,Stacy08}
\f
F^2 (a_N )= \frac{1}{1-e^{-\sqrt{\frac{a_N}{a_0}}}}
\label{function}
\ff
They fit $a_0$ to the data to be
\f
a_0= 1.2 \times 10^{-10}  ms^{-2}
\ff
which is not far from $a^*= c^2 \sqrt{\frac{\Lambda}{3}}$.  Indeed we are not yet fully in a deSitter expansion.

We may note that this has the limits postulated by \cite{MOND}, which are clearly exhibited by the data shown in Figure 1.  For large $\frac{a_N}{a_0}$,  $F \rightarrow 1$, showing that conventional Newtonian
dynamics is restored.  This accords with the observation that dark matter is not needed in the cores of galaxies.
For small  $\frac{a_N}{a_0}$,  $F^2 \rightarrow \sqrt{\frac{a_0}{a_N}}$, giving the MOND formula
\f
a_{obs} = \sqrt{a_N a_0 }
\ff
This gives immediately the baryonic Tully- Fisher relation for the velocity of the flat rotation curves in the exterior of a galaxy,
in terms of its {\it baryonic total mass}, $M$\cite{TF},
\f
v^4 = GM a_0  .
\ff
We may note that the baryonic Tully- Fisher relation is well confirmed\cite{MOND,Sanders,smallscatter}.

Additionally, it must be stressed that Figure 1 shows that the scale $a_0$ is clearly present in the data. Indeed, this scale characterizes the phenomenology of galaxies as it is  a typical scale for spiral galaxies. 

This explanation for the observed mass-descrepancy-acceleration relation (or radial acceleration relation), implies several ramifications by means of which it might be tested.  One is a redshift dependence of $a_0$, corresponding to changes in the cosmic horizon distance and temperature.  Presently there is no evidence for an evolution
of $a_0$ in the Tully-Fisher relation out to redshift of $z=1$\cite{z=1} or $z=1.7$\cite{z=1.7}.

Another is acceleration dependent modifications of the equality of inertial and passive gravitational masses, present universally in processes at very small accelerations of order
$a^*$.  These would have to be tested in space, such as at lagrangian points where the Earth's gravitational acceleration is cancelled\cite{saddle}.

It goes almost without saying that if this proposal has any worth, a goal of research in quantum gravity must be to predict the form of (\ref{function})\footnote{Related ideas have been suggested previously in \cite{MM-related}.}.  We note that the effect in question is based on a renormalization of coupling constants in by averaging over very large scales, and is hence a far infrared effect.  Indeed, there are very good reasons to think that $MOND$ must be due to a new kind of 
non-locality in physics\cite{MM-related,Woodard-nonlocal}.  It is intriguing to wonder if this might have something to do with
relative locality\cite{RL1,RL2}\footnote{The idea that dark energy might be the result of non-locality in loop quantum gravity,
or disordered locality\cite{disordered}, was suggested in \cite{ChandaLee}.  The extension of this to dark matter and MOND was studied in an unpublished 
draft\cite{Leeun}.}.

Of course there remain the challenges to extend MOND to relativistic processes and to address large scale structure formation, the bullet cluster and others.  It is possible  these may be aided by the new point of view proposed here.

\subsection{A second argument}

Here is a second argument for deriving MOND from general relativity, combined with some input from the quantum equivalence principle\footnote{A related argument was proposed in \cite{vP}.}.  We start with the observation of Narnhofer and Thirring\cite{NT} and Deser and Levine\cite{DL1} that an observer in deSitter spacetime, with a steady four acceleration $a$ observes a thermal spectrum with temperature,
\f
T=\frac{\hbar a_5 }{2 \pi c}
\ff
where $a_5$ is the acceleration of the observer's worldline, lifted up into a flat five dimensional embedding of deSitter spacetime.  This is given by (\ref{T1}) with positive $\Lambda$,
\f
a_5 =  \sqrt{a^2 + \frac{c^4 \Lambda }{3 }}= \sqrt{a^2 + \frac{c^4 }{R^2 }}
\label{a5}
\ff
Thus, this acceleration, $a_5$ is the relevant acceleration when we are taking the limit of maximally entangled thermal boosted states with decreasing acceleration to reach the minimally entangled ground state.  Note that when the cosmological constant is non-zero and positive we cannot take this limit all the way, for even observers with vanishing four acceleration, $a$, have
non-vanishing temperature, $T_*$.  So which accelerations are relevant for expressing the equivalence principle in the Newtonian limit?

By construction, the observed kinematical acceleration, $a_k \approx \frac{v^2}{r}$, for $r << R$, is the four acceleration, $a$.  But in what frame is the Newtonian acceleration $a_N^i=- \nabla^i \phi $ applied, and what is its value?  For accelerations
large compared to $a_*$, it doesn't matter as all candidates go to $a$ in the limit, as accords the classical equivalence principle.

But what if we take the proposal that acceleration is tied to temperature seriously, as suggested by \cite{fluctuations} and the quantum equivalence principle?  We then might want to
use $a_5$ for $a_N$ in the non-relativistic limit.  But this makes no sense as it has a positive lower bound, which is $a_*$.
But we must retain that the acceleration vanishes when the force does.  Instead we should choose 
for $a_N$, a function of $a_5$ (and hence of $T$) that goes to zero as $a \rightarrow 0$.  The simplest such function is
\f
\tilde{a}_5 = a_5 - a_*
\ff
Suppose we set $a_N= \tilde{a}_5$?  It follows right away that
\f
a = \sqrt{2a_* a_N}
\ff
which is the MOND relation for small $a_N << a_*$.









\

\section{Conclusions}

We have proposed four principles which a quantum theory of gravity should satisfy.  These together imply the field equations of general relativity, they also express aspects of the $AdS/CFT$ correspondence.   We gave two tentative arguments that suggest that there are quantum gravity effects at very low acceleration, which reproduce the phenomenology of MOND.

I would like to close with an observation and a query.  The observation is that all the cases we have studied situations where variances in time can be neglected, which hence involve static configurations such as uniformly accelerated observers or circular motion.  In these situations there appear applications of equilibrium thermodynamics, at the classical and semiclassical level.  These applications give rise to the Einstein equations, as was proposed in \cite{Ted95}.  But what if we extend our analysis to describe strongly time dependent situations?  Then we  will have to extend our use of thermodynamics to non-equilibrium thermodynamics. 

The query is that, given that we are working in  a context in which equilibrium thermodynamics gives rise to the Einstein equations, which are symmetric under time reversal, will we now see the emergence of a time asymmetric  extension of general relativity, such as are described  in\cite{TA1,TA2}?  

\section*{ACKNOWLEDGEMENTS}

I would like to thank Andrzej Banburski, Jacob Barnett,  Linqing Chen,  Marina Cortes, Bianca Dittrich, Laurent Freidel,  Henriques Gomes,  Andrew Liddle,  Stacy McGaugh,  Mordehai Milgrom, Krishnamohan Parattu, Percy Paul and Vasudev Shyam for very helpful discussions and encouragement.  I am also
indebted to Stacy McGaugh for permission to reproduce Figure 1 and its caption from \cite{MLS}.

This research was supported in part by Perimeter Institute for Theoretical Physics. Research at Perimeter Institute is supported by the Government of Canada through Industry Canada and by the Province of Ontario through the Ministry of Research and Innovation. This research was also partly supported by grants from NSERC, FQXi and the John Templeton Foundation.

\end{document}